\documentclass[english,aps,prb,twocolumn]{revtex4}
\usepackage[T1]{fontenc}
\usepackage[latin1]{inputenc}
\usepackage{graphicx}

\makeatletter

\usepackage{babel}
\makeatother
\begin{document}

\title{Resistivity memory effect in La$_{1-x}$Sr$_{x}$MnO$_{3}$}

\author{E.P.Khlybov $^{\textrm{1,2}}$, R.A.Sadykov$^{\textrm{1}}$, I.J.Kostyleva$^{\textrm{1,2}}$,
W.I.Nizhankovskij$^{\textrm{2}}$, A.J.Zaleski$^{\textrm{3}}$, D.Wlosewicz$^{\textrm{3}}$,
A.W.Giulitin$^{\textrm{1}}$}

\affiliation{$^{\textrm{1}}$Institute of High Pressure, RAS, 142090 Troick, Russia}

\affiliation{$^{\textrm{2}}$International Laboratory of High Magnetic Fields
and Low Temperatures, 53-429 Wroc\l aw, Poland}

\affiliation{$^{\textrm{3}}$Institute of Low Temperature and Structure Research,
PAS, 50-422 Wroc\l aw, Poland}

\begin{abstract}
During the study of magnetoresistivity in La$_{1-x}$Ca$_{x}$MnO$_{3}$
it was found that after cycling of the magnetic field, some kind of
magnetic field memory effect was observed. For La$_{0.5}$Ca$_{0.5}$MnO$_{3}$after
cycling of the magnetic field to 13T and back to zero, {}``frozen''
magnetoresistivity decreases about 20 times comparing to zero field
value, while for La$_{0.47}$Ca$_{0.53}$MnO$_{3}$it is already about
four orders of magnitude. This effect can be observed only for concentration
region $0.45\le x\le0.55$. In zero magnetic field, temperature dependence
of resistivity $\rho\left(T\right)$ shows semiconducting-like behavior,
while after magnetic field cycling it becomes metal-like. So it looks
as we are dealing with magnetic field induced semiconductor (or dielectric)
to metal transition. Such effect can be explained within phase-separation
picture. In zero magnetic field material consists of antiferromagnetic
matrix (insulating phase) and coexisting ferromagnetic, conducting
phase. Magnetic field application causes ferromagnetic phase to form
some kind of conducting channels which shunts semiconducting matrix
phase. Such structure is preserved after reduction of magnetic field,
leaving the material conducting.
\end{abstract}
\maketitle

\section{introduction}

Colossal magnetoresistance, found in perovskite-like (La,M)MnO$_{3}$
(M=Ca, Sr, Ba), is still the subject of intensive studies in many
laboratories around the world \cite{key-1,key-3,key-4}. For the broad
range of their compositions while cooling they undergo magnetic transition
from paramagnetic (PM) into ferromagnetic (FM) state. External magnetic
field increases temperature $T_{c}$ of this transition. Ferromagnetically
ordered phase is highly conducting, so the magnetic transition is
connected with dramatic decrease of resistivity. In parental insulating
compound LaMnO$_{3}$ manganese ions are trivalent. Substitution of
Mn with divalent ions (Sr, Ca or Ba) leads to the appearance of new
compounds with strong ferromagnetism and high conductivity. 

Phase diagram of the La$_{1-x}$Ca$_{x}$MnO$_{3}$ is presented in
the paper by P. Shriffer et al. \cite{key-7}. According to the authors,
for Ca concentrations in the region $0<x<0.15$ the material is insulating
and ferromagnetic; for $0.15<x<0.5$ one deals with conducting, ferromagnetic
(FM) material (for which colossal magnetoresistance effect can be
observed) and for $x>0.5$ compound becomes non conducting and antiferromagnetic
(AF). Investigations of the compounds from the instability region
between AF and FM phases are of particular interest.

\section{Methods}

Samples under study were prepared by solid state diffusion method
of adequately mixed La$_{2}$O$_{3}$, CaCO$_{3}$and MnO$_{2}$.
Mixed powders were fired at 1100$^{o}$C for 5 hours. Such step was
repeated five times with intermediate grindings. Obtained material
was single phased as proved by x-ray diffraction measurements (diffractometer
URD-63 with Fullproof program). From analysis it results that the
obtained materials has Pbnm type structure symmetry. Typical diffractogram
for La$_{0.5}$Ca$_{0.5}$MnO$_{3}$ sample is presented in Fig.1,
\begin{figure}
\includegraphics[%
  scale=0.4]{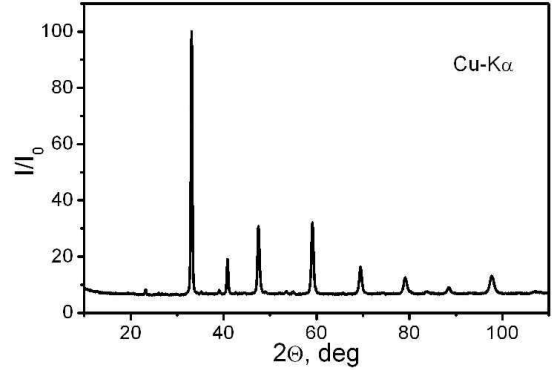}

\caption{Magnetisation on magnetic field dependence for the sample with composition
of La$_{0.5}$Ca$_{0.5}$MnO$_{3}$ at different temperatures: curve
1 -- 4.2K, curve 2 -- 40K, curve 3 -- 200K, curve 4 -- 300K. In the
inset, magnetic field hysteresis is presented at temperature 4.2K.}
\end{figure}
from which following unit-cell parameters were derived: $a=5.436\left(1\right)\textrm{Å}$,
$b=5.422\left(1\right)\textrm{Å}$ and $c=7.638\left(1\right)\textrm{Å}$. 

Magnetic measurements were carried out with use of commercial (Oxford
Instruments) ac susceptometer at the Institute of Low Temperature
and Structure Research (Wroc\l aw, Poland) and specially home-made
capacitive magnetometer at International Laboratory of High Magnetic
Fields and Low Temperatures (Wroc\l aw, Poland). Resistivity measurements
were conducted using standard four-point probe method using Bitter
magnets with magnetic fields up to 14T.

\section{Results and discussion}

Magnetization $\sigma\left(H\right)$ dependence for La$_{0.5}$Sr$_{0.5}$MnO$_{3}$
sample was presented in Fig. 2 %
\begin{figure}
\includegraphics[%
  scale=0.4]{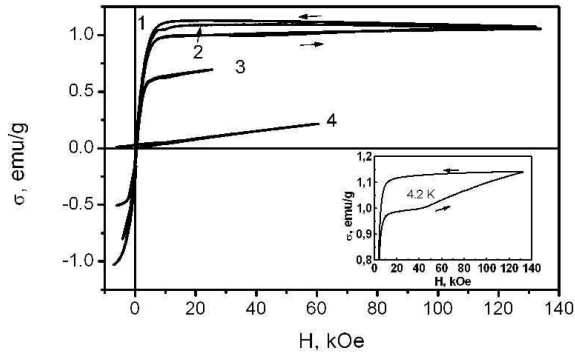}

\caption{Magnetic susceptibility on temperature dependence of La$_{0.5}$Ca$_{0.5}$MnO$_{3}$
at zero magnetic field.}
\end{figure}
and ac susceptibility $\chi\left(T\right)$ in Fig. 3.%
\begin{figure}
\includegraphics[%
  scale=0.4]{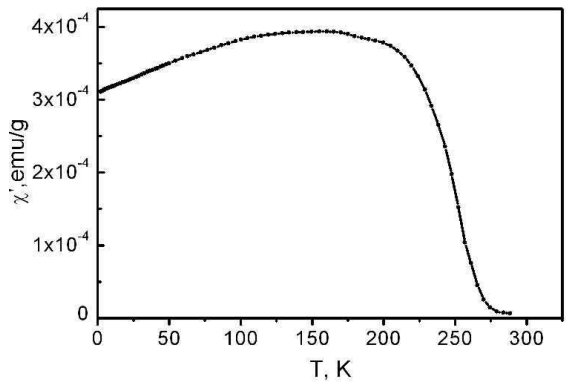}

\caption{Magnetoresistance hysteresis loops at 4.2K after magnetic field cycling
from zero to 5, 8, 10 and 13T respectively. Inset shows hysteresis
of magnetoresistance at 77K. }
\end{figure}
 It is seen that for temperatures below $T_{c}\approx270\textrm{K}$
the dependence is characteristic for ferromagnetically ordered material
but the existence of maximum for about $150\textrm{K}$ may suggest
antiferromagnetic ordering occurrence. It should be pointed out on
characteristic hysteresis loops occurring for high fields on $\sigma\left(H\right)$
dependence (see, inset of Fig. 2). Similar hysteresis loops were observed
by A. Korolev et al. \cite{key-8} while studying La$_{0.9}$Sr$_{0.1}$MnO$_{3}$,
which represents the compound from the transition region between ferromagnetic
and antiferromagnetic phases. The authors explained the appearance
of such hysteresis loops by the existence, apart of ferromagnetic,
also antiferromagnetic phase which can experience metamagnetic transition
during the magnetic field cycling , which influenced the shape of
$\sigma\left(H\right)$ dependence during the increase and decrease
of the magnetic field. 

The question of the coexistence of ordered ferromagnetic and antiferromagnetic
phases in manganites is the part of more broader problem of phase
separation in compounds which consists mixed valency ions (such as
perovskites -- high temperature superconductors and manganites with
colossal magnetoresistance \cite{key-11}\cite{key-12}. Neutron scattering
measurements of La$_{0.5}$Ca$_{0.5}$MnO$_{3}$ \cite{key-13} showed
the existence of transition from paramagnetic to ferromagnetic phase
at $T\cong235\textrm{K}$ followed by appearance of antiferromagnetic
phase at about $T\cong140\textrm{K}$. According to our measurements
it results that ferromagnetic transition takes place at about $T\cong270\textrm{K}$.
It is difficult to determine which of phases plays the role of matrix
and which this of inclusion, both of them coexist and determine the
physical properties of the system.

Until now the colossal magnetoresistance studies showed that the $R\left(H\right)$
dependences are reversible. But our investigation of La$_{1-x}$Ca$_{x}$MnO$_{3}$
($0.45<x<0.55$) revealed new effect - the $R\left(H\right)$ curves
for increasing and decreasing of magnetic field are different; it
looks like the material \char`\"{}remembered\char`\"{} maximal value
of the applied field. 

At Fig. 4%
\begin{figure}
\includegraphics[%
  scale=0.4]{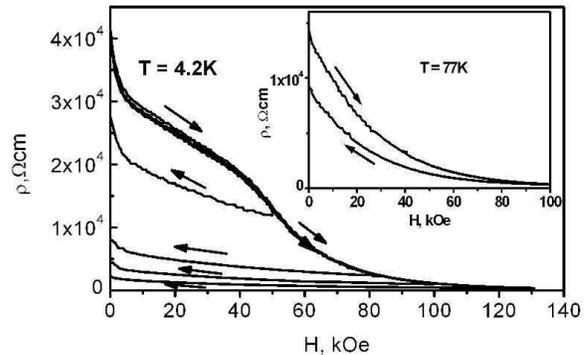}

\caption{Magnetoresistance hysteresis loop at 4.2K for La$_{0.47}$Ca$_{0.53}$MnO$_{3}$.}
\end{figure}
 hysteresis loops of $R\left(H\right)$ dependence are presented for
temperature of 4.2K and different maximal magnetic field values up
to 13T (hysteresis loop for 77K is shown at inset of the Figure 3).
It is seen that during the cycling of magnetic field up to the value
of 13T and back to zero the electrical resistivity decreases about
20 times. The effect of magnetoresistance \char`\"{}memory\char`\"{}
is already visible for so low fields as 100Oe. 

If one cycles the magnetic field value following the procedure : $H=0\rightarrow5\rightarrow2\rightarrow8\rightarrow4\rightarrow13\rightarrow0\textrm{T}$,
it might be seen that each time resistivity increases and decreases
along the same curve. It means that the material behaves like it \char`\"{}remembers\char`\"{}
maximal field it was placed each time. It can be added the observed
effect did not depend on the mutual orientation of the magnetic field
and measuring current.

The observed effect increases with the increasing amount of Ca. For
the compound La$_{0.47}$Ca$_{0.53}$MnO$_{3}$ after the increase
of magnetic field to 14T at 4.2K, resistivity decreases more then
four orders of magnitude (depicted at Fig. 5%
\begin{figure}
\includegraphics[%
  scale=0.4]{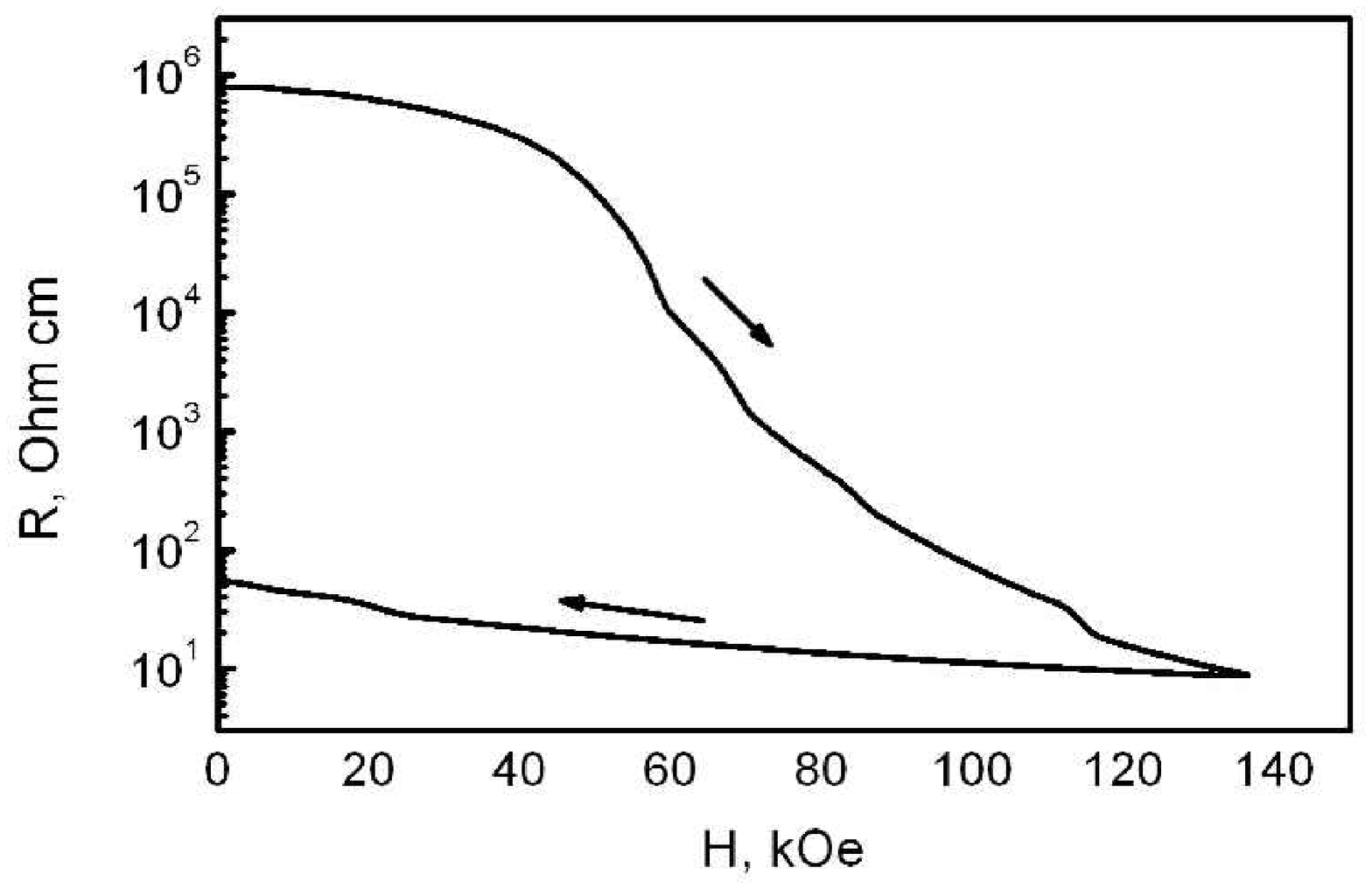}

Temperature dependence of resistivity for La$_{0.5}$Ca$_{0.5}$MnO$_{3}$
without magnetic field - curve 1; at field of 0.2T (curve 2a -- FC,
curve 2b -- ZFC); at 2T (curve 3a -- FC, curve 3b -- ZFC) and 5T (curve
4a -- FC, curve 4b -- ZFC).
\end{figure}
in logarithmic scale). The shape of presented two curves is little
different. For Ca concentration equal to $x=0.5$ characteristic is
sudden drop of resistivity at low magnetic fields, while for the sample
with Ca concentration equal to $x=0.47$ flat resistivity dependence
at low fields is followed by sharp drop for magnetic field of order
of 5T. From practical point of view it looks that more interesting
are materials with \char`\"{}memory\char`\"{} effect existing at low
magnetic fields. For material with composition La$_{0.45}$Ca$_{0.55}$MnO$_{3}$
at 4.2K resistivity is of order of $10^{9}\Omega\times\textrm{cm}$,
but after cycling of magnetic field to 14T it drops to $10^{4}\Omega\times\textrm{cm}$.

Effect of magnetic \char`\"{}memory\char`\"{} may be observed already
for the sample with composition of La$_{0.55}$Ca$_{0.45}$MnO$_{3}$
but its magnitude is rather marginal.

It is interesting to know how long the material will remembers the
value of magnetic field at which it was placed. The time of relaxation
was checked for the sample La$_{0.5}$Ca$_{0.5}$MnO$_{3}$ at the
liquid nitrogen temperature, after magnetic field cycling to 10T.
After reducing of magnetic field to zero the decrease of resistivity
of about 1.5 times was found, comparing to the value before the field
application. During the period of 24 hours frequent resistivity measurements
were conducted. No visible changes of resistivity were noticeable
proving that if there is any relaxation of the magnetic \char`\"{}memory\char`\"{}
effect, its time constant should be very substantial at this temperature.
We managed to measure relaxation of resistivity for the same sample,
but for low temperatures (4.2K) where the difference of resistivity
for no field and after application and reduction of magnetic field
to zero is greatest. But even for this temperature our evaluation
of time constant after application of magnetic field of 14T gives
the relaxation time of order of $10^{1000}$ years (what is more than
the age of our Universe).

Interesting behavior was observed also for temperature dependence
of resistivity R(T). It depends on the manner at which magnetic field
was applied. In Fig. 6 %
\begin{figure}
\includegraphics[%
  scale=0.4]{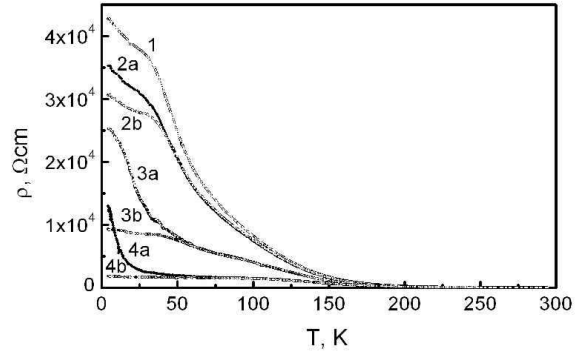}

\caption{Temperature dependence of resistivity for La$_{0.5}$Ca$_{0.5}$MnO$_{3}$
after magnetic field cycling to 0, 5, 8, 10 and 13T and subsequent
decreasing to zero value.}
\end{figure}
we present $R\left(T\right)$ dependence for the sample of composition
La$_{0.5}$Ca$_{0.5}$MnO$_{3}$. Without magnetic field resistivity
shows semiconducting-type behavior and is labeled as 1. If we measure
the same dependence in magnetic field of 0.2T the $R\left(T\right)$
curve has lower values (as might be expected) and we label this curve
as 2a. But after cooling the sample in such field of 0.2T (FC) and
then, without lowering the magnetic field, we again will measure $R\left(T\right)$
dependence, we will have lower resistivity values (we labelled this
curve as 2b). Fig. 6 presents more curves obtained by the same procedure
for different magnetic fields (ZFC and FC).

The results of consecutive measurements of resistivity (after withdrawing
the magnetic field) are presented at Fig. 7.%
\begin{figure}
\includegraphics[%
  scale=0.4]{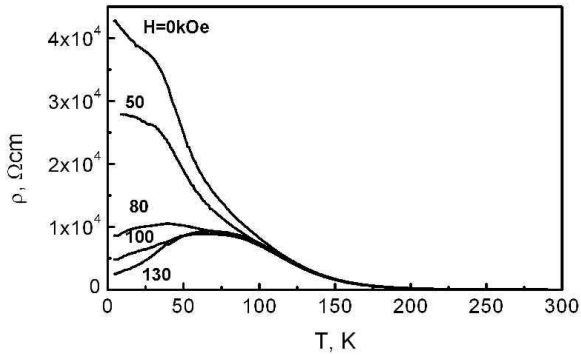}

\caption{Temperature dependence of resistivity of La$_{0.47}$Ca$_{0.53}$MnO$_{3}$
measured at zero magnetic field -- curve 1 and after magnetic field
cycling to 10T -- curve 2.}
\end{figure}
 It is seen that for higher magnetic field values, the low temperature
part of $R\left(T\right)$ dependence shows metal-like behavior rather,
so it looks like there exists semiconductor -- metal transition, which
is induced by the magnetic field application. 

Such a view is supported by the results presented at Fig. 8,%
\begin{figure}
\includegraphics[%
  scale=0.4]{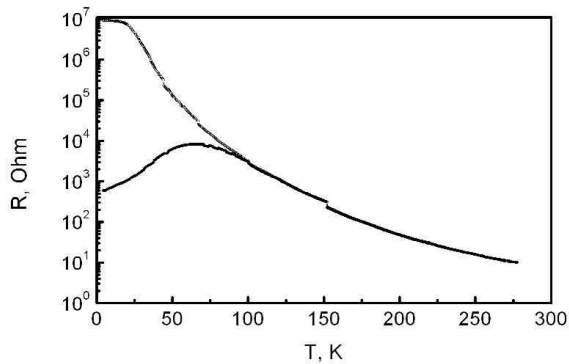}

\caption{Temperature dependence of specific heat for La$_{0.5}$Ca$_{0.5}$MnO$_{3}$.}
\end{figure}
 where temperature dependence of resistivity were presented for La$_{0.47}$Ca$_{0.53}$MnO$_{3}$
compound cooled in zero magnetic field (curve 1) and after magnetic
field cycling $0\rightarrow14\textrm{T}\rightarrow0$ (curve 2). Here
the transition from semiconducting-like to metal-like is much more
pronounced. Similar behavior was already previously reported by F.
Parisi et al. \cite{key-14} for La$_{0.5}$Ca$_{0.5}$MnO$_{3}$
and rather low magnetic field. Their results were also magnetic history
dependent.

At present it is widely accepted that in perovskites (both high temperature
superconductors (HTSC) and manganites with colossal magnetoresistance)
electronic phase separation takes place which leads to stripe structure
existence \cite{key-15}. Such a structure was observed during investigation
of La$_{1-x}$Ca$_{x}$MnO$_{3}$ and La$_{1-x}$Sr$_{x}$MnO$_{3}$
systems at the region between the existence of insulating, antiferromagnetic
and metallic phases. While studying stripe structure at La$_{2-x-y}$Nd$_{y}$Sr$_{x}$CuO$_{4}$
\cite{key-16} it was shown, that if one treats the stripes as elastic
threads, their behavior may be modelled by collective pinning \cite{key-17}. 

The effect of \char`\"{}frozen\char`\"{} magnetoresistivity or the
effect of magnetic field \char`\"{}memory\char`\"{} for La$_{1-x}$Ca$_{x}$MnO$_{3}$
might be explained in the following way. In antiferromagnetic, isolating
matrix of the material coexists high temperature, ferromagnetic phase
which, in the absence of magnetic field, is below the percolation
limit. After application of magnetic field ferromagnetic domains become
connected, giving paths to the electric current flow, and decreasing
the overall resistivity. After removing the applied magnetic field
some kind of pinning-like effect exists, which prevents the separation
of the magnetic and conducting domains. This leads to the effect on
the magnetic field \char`\"{}memory\char`\"{}. 

The physics of such kind of the pinning of magnetic domains is not
very clear yet and needs more, more deep studies. But is supports
in some way the idea of E.L.Nagaev \cite{key-11}, that for colossal
magnetoresistance and high temperature superconductivity effects most
important role is played by electronic phase separation in these materials. 

\begin{acknowledgments}
This work was partially supported by grant RFFI-NNIO no. 01-02-04002.
\end{acknowledgments}

\end{document}